\documentclass[english,a4paper,11pt]{article}

\usepackage{lipsum} 
\usepackage{graphicx} 
\usepackage{amsmath} 
\usepackage{amssymb} 
\usepackage{hyperref} 
\usepackage{bbm}
\usepackage{url}
\usepackage{authblk}
\usepackage[citestyle=authoryear,bibstyle=authortitle,backend=biber]{biblatex}
\usepackage[margin=30mm]{geometry}
\AtEveryBibitem{
    \clearfield{urlyear}
    \clearfield{urlmonth}
    \clearfield{urldate}
}
\newcommand{\abs}[1]{\left|#1\right|}
\newcommand{\indi}{\mathbbm{1}}
\newcommand{\norm}[1]{\left \| #1 \right \|}
\newcommand{\R}{\mathbb{R}}
\newcommand{\N}{\mathbb{N}} 
\newcommand{\KL}[2]{D_{\mathrm{KL}}\left(#1\,\middle\|\,#2\right)}

\begin{document}

\title{Combining Normalizing Flows and Quasi-Monte Carlo}

\author{Charly Andral \\ andral@ceremade.dauphine.fr}

\affil{ CEREMADE, CNRS, Université Paris-Dauphine, Université PSL, 75016 PARIS, FRANCE} 
\date{}

\maketitle

\begin{abstract}
  Recent advances in machine learning have led to the development of new methods for enhancing Monte Carlo methods such as Markov chain Monte Carlo (MCMC) and importance sampling (IS). One such method is normalizing flows, which use a neural network to approximate a distribution by evaluating it pointwise. Normalizing flows have been shown to improve the performance of MCMC and IS. On the other side, (randomized) quasi-Monte Carlo methods are used to perform numerical integration. They replace the random sampling of Monte Carlo by a sequence which cover the hypercube more uniformly, resulting in better convergence rates for the error that plain Monte Carlo. In this work, we combine these two methods by using quasi-Monte Carlo to sample the initial distribution that is transported by the flow. We demonstrate through numerical experiments that this combination can lead to an estimator with significantly lower variance than if the flow was sampled with a classic Monte Carlo.
\end{abstract}

\section{Introduction}

Sampling from a distribution known only by its (unnormalized) density is a problem that has arisen in many different fields such as physics, Bayesian statistics and machine learning. In particular it if often used to estimate an expectation of a function with respect to this distribution, which is known as Monte Carlo integration \parencite{metropolisMonteCarloMethod1949}.  Monte Carlo has been a very important topic in statistics in recent decades \parencite{robertMonteCarloStatistical2010}, especially with the development of Markov Chain Monte Carlo (MCMC),  as the Metropolis-Rosenbluth-Teller-Hastings (MRTH) algorithm \parencite{metropolisEquationStateCalculations1953,hastingsMonteCarloSampling1970} or importance sampling (IS) \parencite{kahnStochasticMonteCarlo1949a,goertzelQuotaSamplingImportance1949}. 

In the recent years, generative methods coming from the machine learning community has been used to enhance existing Monte Carlo methods \parencite{huangAcceleratedMonteCarlo2017,parnoTransportMapAccelerated2018}. In particular, normalizing flows \parencite{tabakDensityEstimationDual2010,tabakFamilyNonparametricDensity2013,dinhNICENonlinearIndependent2015} offer a general framework to approximate the distribution of interest by a measure parametrized by a neural network, while being able to sample from it and compute its density.  The samples from the flow can then be used as the instrumental distribution of an importance sampler \parencite{mullerNeuralImportanceSampling2019,noeBoltzmannGeneratorsSampling2019} or as a proposal for a MRTH kernel \parencite{albergoFlowbasedGenerativeModels2019,gabrieAdaptiveMonteCarlo2022a}. Recently, a comparison of the different ways to use flow to sample was done \parencite{greniouxSamplingApproximateTransport2023}.

On the other hand, (randomized) quasi-Monte Carlo methods \parencite{mcbook} have been for decades to perform numerical integration by replacing the random sample of Monte Carlo by a sequence that covers the hypercube $[0,1]^d$  more uniformly, allowing to obtain better convergence rates than the $O(n^{-1/2})$ of plain Monte Carlo, typically at $O(n^{-1} (\log (n)^d))$ for QMC. Quasi-Monte Carlo can be combined with MCMC \parencite{owenQuasiMonteCarloMetropolis2005,dickDiscrepancyEstimatesVariance2014,dickDiscrepancyBoundsUniformly2016,liuLangevinQuasiMonteCarlo2023} or with importance sampling \parencite{dickWeightedDiscrepancyBound2019,keAdaptiveImportanceSampling2023}.

As far as we are aware of, this combination of normalizing flows with quasi-Monte Carlo has only been tried in \parencite{wenzelQuasiMonteCarloFlows2018}. The scope of the paper was different, as the authors were interested in QMC to minimize the error of the stochastic gradient for the training of the flow, even though they mentioned that the QMC sample could be used to reduce the variance when estimating an expectation. In particular, the combination with IS and MCMC was not studied.

In this work, we explore the combination of using normalizing flows as an approximation of the target distribution and using quasi-Monte Carlo to sample the initial distribution of the flow. We will show that in same cases, this combination lead to a estimator with a significantly lower variance than if the flow was sampled from i.i.d variables.

\section{Normalizing Flows}
\label{seq:nf}
\subsection{Using an instrumental distribution for Monte Carlo}
\label{seq:montecarlo}

We place ourselves in the classic Monte Carlo setting, where, for a given distribution $\pi$ on $\mathbb{R}^d$, we want to estimate the expectation of a test function $f$ with respect to $\pi$, defined as

\begin{equation*}
    \pi(f) :=  \mathbb{E}_\pi \left[ f(X)\right] =  \int_{ \mathbb{R}^d} f(x) \pi(dx)
\end{equation*}
in the case where a direct sample from $\pi$ is not available and the integral is intractable. 

\paragraph{Importance Sampling} One solution, if we can sample from an instrumental distribution $\rho$ close enough to $\pi$,  is to use the importance sampling estimator by noticing that:

\begin{equation*}
    \pi(f) = \int_{ \mathbb{R}^d} f(x) \frac{\pi(x)}{\rho(x)} \rho(dx) = \mathbb{E}_\rho \left[ f(X) \frac{\pi(X)}{\rho(X)} \right].
\end{equation*}
Therefore, if we can sample from $\rho$, we can estimate $\pi(f)$ by:

\begin{equation}
    \pi(f) \approx  \hat I_{\text{IS}}(f) := \frac{1}{N} \sum_{i=1}^N f(X_i) w(X_i),
    \label{eq:is:plain}
\end{equation}
for $X_i \sim \rho$ and $w(x) = \frac{\pi(x)}{\rho(x)}$, which is called the importance weight. 

If the densities $\pi$ and/or $\rho$ are known only up to a normalizing constant, the weight $w(x)$ is known up to a constant. We denote the unnormalized importance weight by $\tilde w$ := $c \cdot w $ for an unknown $c \in \mathbb{R}$. In that case, the estimator $\hat I_{\text{IS}}(f)$ cannot be used as is and is replace by the self-normalized importance sampling estimator:

\begin{equation}
  \hat I_{\text{SNIS}} = \frac{\sum_{i=1}^{n}\tilde  w(X_i)f(X_i)}{\sum_{i=1}^{n} \tilde w(X_i)}.
  \label{eq:snis:plain}
\end{equation}

While $\hat I_{\text{IS}}$ is unbiased (i.e. $\mathbb{E}[\hat I_{\text{IS}}] = \pi(f)$), $\hat I_{\text{SNIS}}$ is biased due to the ratio of the two estimates.

\paragraph{Independent MRTH algorithm}
Another solution is to use a Metropolis-Rosenbluth-Teller-Hastings  (MRTH) kernel using $Q(x,dy) = \rho(dy)$ as the proposal kernel, which we call the independent MRTH (iMRTH) kernel. This produces a Markov chain that admits $\pi$ as its invariant distribution. More precisely, at a given time $t$ and point $X_t$, the new state $X_{t+1}$ is defined by

\begin{equation*}
  X_{t+1} = \begin{cases} Y_{t+1} & \text{with probability } \alpha(X_t,Y_{t+1}) \\ X_t & \text{otherwise }  \end{cases},
\end{equation*}

where $Y_{t+1}$ is drawn from $\rho$ and $\alpha(x,y) := \min\left(1, \frac{\pi(y)\rho(x)}{\rho(y)\pi(x)}\right)$ is called the acceptance rate.

\paragraph{Importance Markov Chain}

Recently \cite{andralImportanceMarkovChain2023a} proposed a new method called Importance Markov Chain (IMC) that is in between  importance sampling and MCMC. In its independent version (iIMC), it corresponds to drawing a sample $X_i \sim \rho$ and $N_i \in \N \sim R(X_i,.)$ for $R$ a kernel such that $\mathbb{E}\left[N_i \middle| X_i\right]= \kappa \cdot w(X_i)$.
The parameter $\kappa$ is a fixed constant and $w$ the importance weight function. The final chain is obtained by concatenating over $i \in 1:n$ the $X_i$ replicated $N_i$ times.
\subsection{Normalizing Flows}

Normalizing flows are a type of model that can transform a simple distribution into a more complex one using a neural network trained to target a distribution through pointwise evaluation.

Formally, a normalizing flow is an invertible and differentiable map $T$ from $\mathbb{R}^d$ to $\mathbb{R}^d$ . Given a reference probability measure $\mu$ (often a standard Gaussian) we can define a new probability measure $\nu$ as the pushforward of $\mu$ by $T$, i.e. $\nu = T_\sharp \mu = \mu (T^{-1}(.))$. The flow is trained by minimizing the Kullback-Leibler divergence between $\nu$ and the target distribution $\pi$. A sample from the flow can be generated by sampling  $X \sim \mu$ and then taking $T(X)\sim \nu$.

The density of the pushforward measure $\nu$ can be computed using the change of variable formula: 
\begin{equation*}
    \nu(x) = \mu(T^{-1}(x)) \abs{\det(J_{T^{-1}}(x))},
\end{equation*}

where $J_{T^{-1}}$ is the Jacobian of $T^{-1}$, the inverse map of $T$.

Normalizing flows are designed to satisfy several useful properties: the inverse map $T^{-1}$ and the determinant of its Jacobian should be easily evaluated to allow for the computation of the density of $\nu$; $T$ should be easily computable to efficiently generate samples from $\nu$; and the class of flows should be flexible enough to approximate any distribution. For a more detailed introduction of normalizing flows and the different models, we refer to  \parencite{kobyzevNormalizingFlowsIntroduction2021,papamakariosNormalizingFlowsProbabilistic2021a}.

In our numerical experiments, we will use two different types of flows: the real-valued non-volume preserving (RealNVP) transforms of \parencite{dinhDensityEstimationUsing2017} and the rational quadratic neural spline flows (RQ-NSF, in their coupling form) of \parencite{durkanNeuralSplineFlows2019a}, with the latter being more flexible than the former.

\paragraph{Coupling flows}
Coupling flows, introduced by \cite{dinhNICENonlinearIndependent2015}, are used in RealNVP and RQ-NSF. Each layer of the flow is defined by a function $f$ that takes an input vector $x  = (x_i) \in \R^d$, a $d_1<d$, and defines the output $y = f(x)$ elementwise as follows:

\begin{equation}
  f(x_i) = \begin{cases} x_i & \text{if } i \leq d_1 \\ g(x_i,\theta(x_{1:d_1}))& \text{otherwise} \end{cases}.
  \label{eq:coupling}
\end{equation}
Here,  $g$ is a certain function with a parameter $\theta$ that depends only on  $x_{1:d_1} := (x_1,\dots,x_{d_1})$. For RealNVP, $g$ is an affine function; for RQ-NSF, $g$ is a rational quadratic spline \parencite{gregoryPiecewiseRationalQuadratic1982a}. The parameter $\theta$ is called the conditioner and is usually a neural network (multilayer perceptron) with $x_{1:d_1}$ as input. In practice, $d_1 = d/2$ and the function uses a mask to alternate the dimensions that are kept unchanged and the ones that are transformed.

If $g$ is invertible, $f$ is invertible too, and the inverse of $f$ is given by:

\begin{equation*}
  f^{-1}(y_i) = \begin{cases} y_i & \text{if } i \leq d_1 \\ g^{-1}(y_i,\theta(y_{1:d_1}))& \text{otherwise} \end{cases}.
\end{equation*}
By construction, the Jacobian of $f$ (and $f^{-1}$) is triangular, and therefore its determinant is the product of the diagonal elements. 

For a more general definition of coupling flows where $f$ is not applied elementwise and for the defintion of autoregressive flows (not used in this article), we refer to \parencite{kobyzevNormalizingFlowsIntroduction2021}.

\paragraph{Training of the flow}

In most of the cases (e.g. for RealNVP and RQ-NSF), the flow is parametrized by a neural network. Let us write $T_\theta$ for $\theta \in \mathbb{R}^d$  and $\nu_{\theta} := (T_\theta)_\sharp \mu $ to explicit the dependence of the flow on the parameter $\theta$. The goal of the training in to find $\theta $ such that $\nu_{\theta}$ is close to $\pi$. 

This proximity is usually measured by the (forward) Kullback-Leibler (KL) divergence between the target distribution $\pi$ and the distribution of the flow $\nu_{\theta}$:

\begin{equation}
   \KL{\mu}{\nu_\theta} := \int_{\mathbb{R}^d} \log \left(\frac{\mu(x)}{\nu_\theta(x)}\right) \mu(x) dx. 
  \label{eq:kl:forward}
\end{equation}

In some cases, the reverse KL divergence $\KL{\mu}{\nu_{\theta}}$ is minimized instead. For more details on the differences between the forward and reverse KL divergences see \parencite{papamakariosNormalizingFlowsProbabilistic2021a}.

This paper does not propose a new training for the flow. Therefore we will use different trainings from other works in our numerical experiments. 
\section{(Randomized) Quasi-Monte Carlo}
\label{seq:qmc}

\subsection{Quasi-Monte Carlo}
Quasi-Monte Carlo (QMC) methods are used to solve the problem of integrating a function over the unit hypercube $[0,1]^d$. Unlike (plain) Monte Carlo methods, QMC methods use a deterministic sequence of points that is designed to cover more uniformly the space, rather than random points. For an introduction to QMC, see \parencite{mcbook}.  The uniformity of the sequence is measured by the star-discrepancy, defined fora set of points $x_1, \dots, x_n \in [0,1]^d$ by:

\begin{equation*}
    D^\star_n(x_1,\dots, x_n) = \sup_{\mathbf{a} \in [0,1]^d} \abs{\frac{1}{n}  \sum_{i=1}^n \indi_{x_i \in [0,\mathbf{a})} - \prod_{i=1}^d a_i},
\end{equation*}
where $\mathbf{a} = (a_1,\dots,a_d)$ and $[0,\mathbf{a}) = \prod_{i=1}^d[0,a_i) \subseteq [0,1]^d$.

The error of the QMC estimator built with $x_1,\dots, x_n$ is controlled by the Koksma-Hlawka inequality \parencite{koksma1942een,hlawka1961funktionen}:
\begin{equation*}
    \abs{\frac{1}{n} \sum_{i=1}^{n} f(x_i)  - \int_{[0,1]^d} f(x)dx } \leq D^\star_n(x_1,\dots, x_n) V_{\text{HK}}(f),
\end{equation*}
where $V_{\text{HK}}(f)$ is the total variation of $f$ in the sense of Hardy and Krause. For $d=1$, it coincides with the classic total variation. For the general definition, see \parencite{mcbook}.

A QMC sequence is constructed such that $D^\star_n (x_1, \dots, x_n) = O(n^{-1} (\log n)^d)$. Therefore, if $ V_{\text{HK}}(f) < \infty$, the QMC estimator has a convergence rate of $O(n^{-1} (\log n)^d)$, which is better than the $O(n^{-1/2})$ of plain Monte Carlo. There are two types of QMC methods: lattice rules and sequences. 

\paragraph{Lattice rules}  Lattice rules are the first type of QMC methods. They involve taking a generating vector $z \in \mathbb{N}^d$ and a fixed $n$, and defining a rank-1 lattice by:

\begin{equation*}
  x_i = \frac{(i-1) z}{n} \mod 1,
\end{equation*}
for $i \in 1:n$, where the modulo is taken componentwise. The generating vector $z$ must be carefully chosen to have a low discrepancy. To be efficient, exactly $n$ points must be used for the estimate, and more points are needed, a new lattice must be computed. Lattice rules are particularly used in numerical analysis, but not much in statistics. We will in this paper focus on the second type of QMC methods, the sequences and digital nets.

\paragraph{QMC sequences }
The second type of QMC methods are sequences, such as the Halton sequence or the Sobol' sequence. \parencite{halton1960efficiency,sobol1967distribution}.  The latter is part of what is called digital nets. Unlike lattice rules, sequences are infinite, and more points can be added to the sequence without having to recompute the whole sequence. However, due to the construction of the Sobol' sequence, to keep the balance of the sequence, it is highly recommended to use a number of points $n$ that is a power of 2, and not to drop the first point \parencite{owenDroppingFirstSobol2022}

\subsection{Randomized Quasi-Monte Carlo}

As a QMC sequence is deterministic, it does not allow for the estimation of the error of integration. Therefore, a variant of quasi-Monte Carlo called randomized quasi-Monte Carlo (RQMC) was developed. The sequence $x_1, \dots, x_n$ is no longer deterministic, but still has an almost surely $O(n^{-1} (\log n)^d)$ star-discrepancy while having every marginal distribution uniform on $[0,1]^d$. The latter property also implies that the estimator is unbiased, while the plain QMC estimator is usually biased. For instance, one can take a lattice rule $x_1,\dots, x_n$and apply a common shift $U \sim \mathcal{U}([0,1]^d)$ to all the points: $\tilde x_i = x_i + U$ \parencite{cranleyRandomizationNumberTheoretic1976}. 

For sequences like the Halton or Sobol' sequences, the randomization is different. Instead of a global shift, the randomization is done by scrambling the sequence. In a nutshell, scrambling can be describe as a permutation of the digits (in a certain base) of the points. The permutations are the same for all the points. Specific scrambling was developpd for the Halton sequence \parencite{owenRandomizedHaltonAlgorithm2017a} and the Sobol' sequence \parencite{owen1995,matousekL2DiscrepancyAnchoredBoxes1998}.

In the case of scrambled Sobol' sequence, the randomization can improve the error from $O(n^{-1} (\log n)^{d-1})$ to $O(n^{-3/2}(\log n)^{(d-1)/2})$ \parencite{owenScrambledNetVariance1997}. Running RQMC instead of QMC allows for the estimation of the error by repeating the experiment and even computing confidence intervals \parencite{lecuyer:hal-04088085}.

\section{Quasi-Monte Carlo for Normalizing Flows}

\subsection{From $[0,1]^d$ to $\mathbb{R}^d$ }
\label{sec:qmc:transform}
In some cases, an integral on $\mathbb{R}^d$ can be transformed via a map as an integral on  $[0,1]^d$. For instance, a multidimensional Gaussian random variable can be written as a transform $S:  [0,1]^d \to \mathbb{R}^d$  of a uniform random variable on the hypercube. This can be done by taking the inverse transform or the Box-Muller transform \parencite{box1958note}. This allows for the use of QMC.

The inverse transform is defined as follows: if $U_1 \sim \mathcal{U}([0,1])$ and $F$ is the cumulative distribution function of a standard Gaussian, then $ F^{-1}(U_1) \sim \mathcal{N}(0,1)$. A multidimensional standard Gaussian random variable can be obtained by applying this transform componentwise to $U \in \mathcal{U}([0,1]^d)$. 

The Box-Muller transform is defined as follows: if $U_1, U_2 \sim \mathcal{U}([0,1])$, then $X_1 =  \sqrt{-2 \log(U_1)} \cos(2\pi U_2)$ and $X_2 =  \sqrt{-2 \log(U_1)}\sin(2\pi U_2)$ are two independent standard Gaussian variables. This can be extended in higher dimension by taking $U \in \mathcal{U}([0,1]^d)$ and applying the transform by blocks of size 2.

The (R)QMC sample $x_1,\dots, x_n$ is now transformed to the Gaussian RQMC sample $S(x_1), \dots, S(x_n)$.  It is worth noting that in that case, as this transform is not bounded, $V_{\text{HK}}(S)= \infty$ and the Koksma-Hlawka inequality is useless. However, RQMC can still show better convergence rate than plain MC (and QMC) \parencite{owenHaltonSequencesAvoid2006}.

\subsection{Importance sampling with NF and QMC}

First, let us rewrite $\pi(f)$ as an integral over the hypercube $[0,1]^d$. From now on, the instrumental distribution for the IS is set to be the measure of the flow $\rho$. Recall that if $X \sim \mathcal{N}(0,I_d)$, then $T(X)  \sim \rho $. Denote S as the mapping from $[0,1]^d$ to $\mathbb{R}^d$ such that if $U \sim \mathcal{U}([0,1]^d)$, then $S(U) \sim \mathcal{N}(0,1)$. Then, by composing the two maps, we have that $(T \circ S)(U) \sim \rho $. We can finally write $\pi(f)$ as an integral on the hypercube:

\begin{equation*}
    \pi(f) = \int_{[0,1]^d} \left(f\cdot  w\right)\left((T \circ S)(u)\right) d u 
\end{equation*}
where $w(x) = \pi(x) / \rho(x)$ the importance weight. 

Taking $x_1, \dots,  x_n$ from a RQMC sequence, we get the following estimator:

\begin{equation}
    \hat I^{\text{RQMC}}_{\text{IS}}(f) = \frac{1}{n} \sum_{i=1}^n \left( f \cdot w\right)\left((T \circ S)(x_i)\right).
    \label{eq:is:qmc}
\end{equation}

If the densities are not normalized, the (self-normalized) estimator is :
\begin{equation}
  \hat I^{\text{RQMC}}_{\text{SNIS}}(f) = \frac{ \sum_{i=1}^n \left( f \cdot \tilde w\right)\left((T \circ S)(x_i)\right)}{\sum_{i=1}^{n} \tilde w((T \circ S)(x_i))}
  \label{eq:snis:qmc}
\end{equation}

As marginally $x_i \sim \mathcal{U}(0,1)$ for all $i \in 1:n$, $\hat I^{\text{RQMC}}_{\text{IS}}(f)$ is an unbiased estimator of $\pi(f)$ and $\hat I^{\text{RQMC}}_{\text{SNIS}}(f)$ is biased.

Using the exact same notation, but using a classic Monte Carlo sample $x_1, \dots, x_n$ i.i.d. and uniformly distributed on $[0,1]^d$, we can define the two associated Monte Carlo estimators $\hat I^{\text{MC}}_{\text{IS}}(f)$ and $\hat I^{\text{MC}}_{\text{SNIS}}(f)$. They coincide with the estimator from \eqref{eq:is:qmc} and \eqref{eq:snis:qmc} if we do the change of variables. 

The case of importance sampling using QMC and Gaussian instrumental distribution was recently studied in \parencite{heErrorRateImportance2023} and it can achieve a  $O(n^{-1 + \varepsilon})$ converge rate in some cases. A study of the star-discrepancy in the specific case of importance sampling was conducted in \parencite{dickWeightedDiscrepancyBound2019}.
\section{Numerical experiments}

For the different experiments, we will use two flows from the literature and show how using quasi-Monte Carlo instead of Monte Carlo can lead to a use variance reduction. In all the experiments, we will use the same number of points for the RQMC and the MC estimator. The RQMC sample are computed using the SciPy package in Python \parencite{royQuasiMonteCarloMethods2023}.  The implementation for the Sobol' sequence use the work of \parencite{joeConstructingSobolSequences2008} and the scrambling of \parencite{matousekL2DiscrepancyAnchoredBoxes1998}. The implementation for the Halton sequence follows \parencite{owenRandomizedHaltonAlgorithm2017a}. No pure QMC was used; every QMC mentioned in the figures is RQMC. The code for the experiments is available at \url{https://github.com/charlyandral/QMC_norm_flows}.
\subsection{Mixture of Gaussians in 2D}

\subsubsection{Setting}
For this first experiment, we take the mixture of Gaussians example of \parencite{midgley2023flow}, following the code available at \url{https://github.com/lollcat/fab-torch}. The target density $\pi$ is mixture of 40 Gaussians on $\R^2$. The training is done using an $\alpha$-divergence for $\alpha=2$:
\begin{equation*}
  D_2(\pi\| \nu_\theta) := \int \frac{\pi(x)^2}{\nu_\theta(x)}dx = \mathbb{E}_g\left[\frac{\pi(X)^2}{\nu_{\theta}(X)g(X)}\right]
\end{equation*}
where $g$ is an importance function. They took $g$ to be an approximation of $\pi^2/\nu_\theta$ using annealed importance sampling. The flow is a realNVP from \parencite{dinhDensityEstimationUsing2017} with 15 layers. 

In this example, as the normalizing constant of the target is known, the (non-self-normalized) importance sampling estimator can be use as defined by \eqref{eq:is:qmc}.

For a same flow, we compare the value of the importance sampling estimator using a classic Monte Carlo sample and a scrambled Sobol' sequence with an inverse map. A total of $2^{16}$ points are taken for every estimate. This is repeated 200 times to get estimate of the variance and the boxplot. Seven different test functions are evaluated : $\phi_1 :x \mapsto x_1 $, $\phi_2 :x \mapsto x_2$, $\phi_3 :x \mapsto x_1^2$, $\phi_4 :x \mapsto x_2^6$, $\phi_5 : x: x_1 x_2$, $\phi_6 :x \mapsto \sin(x_1) \cos(-x_2 /10)$,  $\phi_7  = \mathbf{1}_{x_1>30}$. 

\subsubsection{Results}

\begin{figure}[h]
  \vspace{.3in}
  \centerline{\includegraphics[width=1.\linewidth]{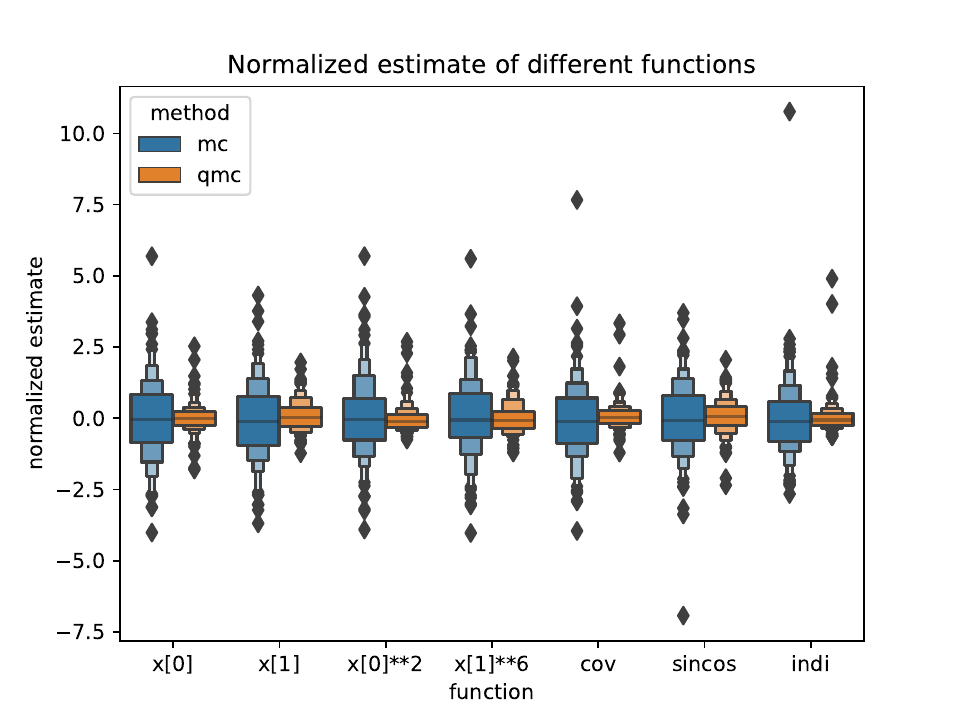}}
  \vspace{.3in}
  \caption{Comparison of the importance sampling estimator using an i.i.d. sample and a Sobol' sequence for different test functions for the Gaussian mixture.}
  \label{fig:gmm:normalized}
  \end{figure}

The results can be visualized in the boxplot of Figure \ref{fig:gmm:normalized}, which shows the different values of $\hat I^{\text{RQMC}}_{\text{IS}}(\phi_i)$ and $\hat I^{\text{MC}}_{\text{IS}}(\phi_i)$. For the sake of readability, the values of the estimator are normalized  for every function,  i.e. the mean of the estimates is first subtracted and then the result is divided by the standard deviation. As the this linear scaling is the same for the two estimators (for a fixed $\phi_i$), the boxplots are still comparable.

The ratio of the empirical standard deviations of the two estimators 
\begin{equation}
  r_i =  \frac{\hat \sigma \left(\hat I^{\text{MC}}_{\text{IS}}(\phi_i)\right)}{\hat \sigma \left(\hat I^{\text{RQMC}}_{\text{IS}}(\phi_i)\right)}
  \label{eq:ratio:var:est}
\end{equation}
 is given in Table \ref{table:gmm:ratio:var}.

  \begin{table}[h]
    \caption{Ratio of the standard deviations between RQMC and MC estimates for several test functions for the gaussian mixture.} \label{table:gmm:ratio:var}
    \begin{center}
      \begin{tabular}{rrrrrrr}
        \hline
        $\phi_1$ & $\phi_2$ & $\phi_3$ & $\phi_4$ & $\phi_5$ &$\phi_6$ & $\phi_7$ \\
        \hline
        2.829 & 2.285 & 2.182 & 2.787 & 2.757 & 2.465 & 2.374 \\
        \end{tabular}
    \end{center}
    \end{table}

The results for the seven different test functions are quite similar. They show a significant variance reduction when using a scrambled Sobol' sequence instead of a classic Monte Carlo sample. The ratio of the standard deviations $r_i$ is quite consistent between the different functions, going from $2.182$ to $2.829$.

\subsection{Dualmoon target}

\subsubsection{Setting}

For the second experiment, we will follow the methodology of \parencite{gabrieAdaptiveMonteCarlo2022a}  to train the flow. This is implemented in the package FlowMC in Python, \parencite{wongFlowMCNormalizingflowEnhanced2022}. The target we used is called the dualmoon in this package.  In this case, the authors choose the forward KL divergence  \eqref{eq:kl:forward} as the training objective. As it requires computing an expectation with respect to the target, direct simulation in not possible, and the optimization of the flow is done concurrently with an adaptive MCMC. 

The target distribution $\pi$ is defined, for a dimension $d$ that can change, by:

\begin{align*}
  \pi(x) \propto &\exp \bigg \{  -\frac{1}{2}\left(\frac{\norm{x}-2}{0.1}\right)^2 + \\
  &\sum_{i=1}^{d} \log \left( e^{-\frac{1}{2}(\frac{x_i+3}{0.6})^2} + e^{-\frac{1}{2}(\frac{x_i-3}{0.6})^2}\right)\bigg\}.
\end{align*}

The density is known up to a normalizing constant, and the self-normalized importance sampling estimator \eqref{eq:snis:qmc} is therefore used for the rest of this section. The target presents multimodality as the number of modes is $2^d$. 

The flows are  the rational quadratic neural spline flows (RQ-NSF) from \parencite{durkanNeuralSplineFlows2019a}, in their masked coupling form. The number of layers and bins will change in the different illustrations. Except of section \ref{sec:effect:sample}, the RQMC sequence is always a scrambled Sobol' sequence with a inverse transform.

\subsubsection{Effects of the dimension}

In this first experiment, we study the effect of the dimension on the variance of the estimator. We will take $d=2,3,4,5,6,7,8,9,10$ and for each dimension, we train 5 flows with 6 layers and 8 bins and a $(32,32)$ perceptron for the hidden layers. For each dimension, we take 50 samples of $2^{16}$ points from scrambled Sobol' sequences and classic Monte Carlo samples. Two test functions are evaluated: $\phi_1 : x \mapsto x_1$ and $\phi_2 : x \mapsto \sin(10 \cdot x_1) \cdot x_1^4$. Note that this time, as the target $\pi$ is symmetric and the two test functions are odd, the true value to estimate is $ \pi(\phi_1) = \pi(\phi_2) = 0$. This allows computing the ratio if the absolute errors between the two estimators. Table \ref{table:dualmoon:dimension:std} shows the ratio of the standard deviations between the two estimators and for the two test functions, Table \ref{table:dualmoon:dimension:error} shows the ratio of the absolute errors.

\begin{table}[h]
  
  \begin{center}
    \begin{tabular}{lrrrrrrrrr}
      \hline
      dim & 2 & 3 & 4 & 5 & 6 & 7 & 8 & 9 & 10 \\
      \hline
      $\phi_1$ & 15.06 & 4.60 & 1.88 & 1.99 & 1.73 & 1.68 & 1.48 & 1.59 & 1.37 \\
      $\phi_2$ & 39.03 & 6.31 & 2.71 & 2.49 & 2.14 & 1.85 & 1.85 & 1.69 & 1.71 \\
      
      \end{tabular}
      \caption{Effect of the dimension on the ratio of the standard deviations between RQMC and MC estimates for the dualmoon target.} \label{table:dualmoon:dimension:std}
  \end{center}
  \end{table}

  \begin{table}[h]

    \begin{center}
      \begin{tabular}{lrrrrrrrrr}
        \hline
        dim & 2 & 3 & 4 & 5 & 6 & 7 & 8 & 9 & 10 \\
        \hline
        $\phi_1$ & 20.04 & 4.80 & 2.02 & 2.00 & 1.68 & 1.63 & 1.51 & 1.59 & 1.39 \\
        $\phi_2$  & 48.31 & 6.42 & 2.79 & 2.61 & 2.21 & 1.88 & 1.92 & 1.69 & 1.72 \\
        
        \end{tabular}
        \caption{Effect of the dimension on the ratio of the absolute errors between RQMC and MC estimates for the dualmoon target.} \label{table:dualmoon:dimension:error}
    \end{center}
    \end{table}

The results in Tables \ref{table:dualmoon:dimension:std} and \ref{table:dualmoon:dimension:error} show very similar results: the efficiency of RQMC over MC is decreasing with the dimension and  also depends on the test function. The test function $\phi_1$ is more simple than $\phi_2$, and thus the variance reduction is more significant. In dimension 2, the improvement is substantial, with a ratio of standard deviations up to $39.03$ and up to $48.31$ for the ratio of errors. For higher dimensions, the two types of ratios shows very similar results, less important than in dimension 2 but still significant. As RQMC has better convergence rate that MC, the ratio is expected to increase with the number of samples. This will be detailed in section \ref{sec:effect:sample}.

This reduction of efficiency with the dimension is not surprising as the quality of the QMC estimate highly depends on the regularity of the function to integrate, and the flow becomes more complex with the dimension. In particular, RQMC is sensitive to effective dimension \parencite[chapter 17.2]{mcbook}. Intuitively, the effective dimension is a way to measure the complexity of the function to integrate. The effective dimension if low if the function can be written a sum of functions depending on  a small number of inputs.  An RQMC integration will perform better compared to MC if the effective dimension if small, even if the dimension of the space $d$ is large.  If the two test functions $\phi_i$ seems to have an effective dimension of one, the true function to integrate with RQMC is $\left( \phi_i\cdot  w\right)\left((T \circ S) \right)$. The flow $T$ defined by a coupling of equation \ref{eq:coupling} is designed to make the entries interact. Therefore, the effective dimension of the true function to integrate is higher than the effective dimension of $\phi_i$.

\subsubsection{Effect of the complexity of the flow}
\begin{figure}[h]
  \vspace{.25in}
  \centerline{\includegraphics[width=1.1\linewidth]{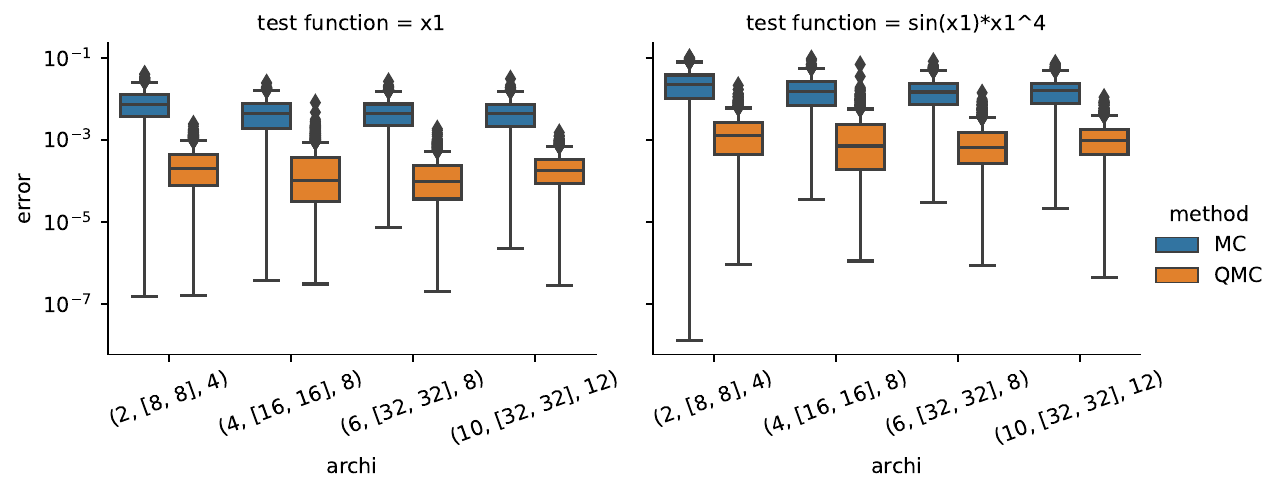}}
  \vspace{.3in}
  \caption{Impact of the complexity of the flow on absolute error between RQMC and MC estimates for the dualmoon target in dimension 2 }
  \label{fig:dualmoon:archi}
  \end{figure}

We compare the effect of the complexity of the flow on the estimation. For a fixed dimension $d$, we train five flows in each of the setting form the parameters, written as (number of layers, hidden size of the conditioner, number of bins): (2,[8,8],4), (4,[16,16],8), (6,[32,32],8) and (10,[32,32],12). All flows have the same number of training loops. For each flow, we take 50 samples of $2^{16}$ points from scrambled Sobol' sequences and classic Monte Carlo samples. We evaluate two test functions: $\phi_1 : x \mapsto x_1$ and $\phi_2 : x \mapsto \sin(10 \cdot x_1) \cdot x_1^4$.  The results are shown in Figure \ref{fig:dualmoon:archi}. 

Consistent with the previous section, the variance reduction is more significant for the simple test function $\phi_1$ than for $\phi_2$. Although the simplest flow has a slightly higher average error, the other three flows have similar results. The most performing is the (6,[32,32],8) flow. The results obtained here are consistent with the average training loss of the four flows, respectively 1.43, 0.50, 0.42, 0,43. The ratio of RQMC error over MC is similar is the four cases.

\subsubsection{Effect of the choice of the RQMC sequence}
\label{sec:effect:sample}
\begin{figure}[h]
  \vspace{.25in}
  \centerline{\includegraphics[width=1.1\linewidth]{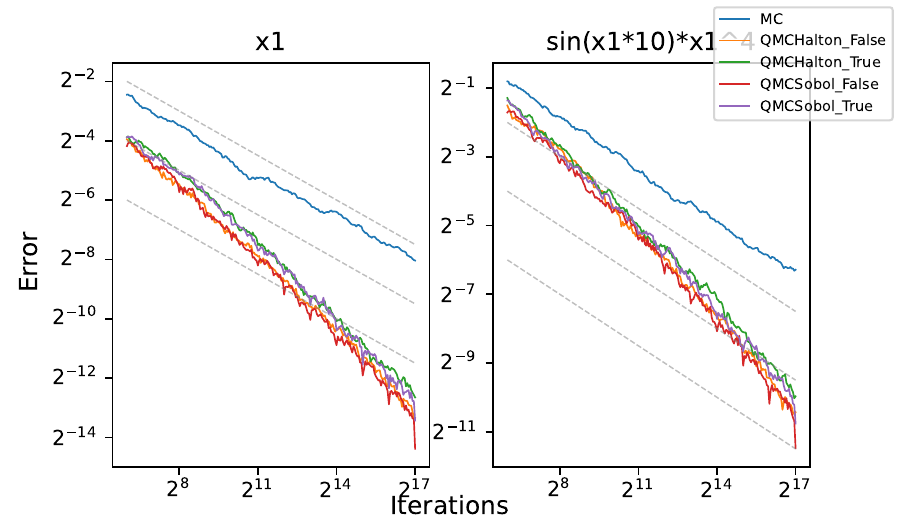}}
  \vspace{.3in}
  \caption{Convergence of the error of the importance sampling estimator for the dualmoon target in dimension 2 in log-log scale. True means inverse transform, false means Box-Muller. Dotted line corresponds to the convergence rate of $O(n^{-1/2})$ of plain Monte Carlo.}
  \label{fig:dualmoon:seq}
  \end{figure}

In this section, we compare Sobol' and Halton sequences, as well as two mappings from $\mathcal{U}((0,1)^d)$ to $\mathcal{N}(0,I_d)$, the inverse and the Box-Muller transform described in section \ref{sec:qmc:transform}. This gives a total of four methods. The dimension is set to $d=2$ as it was showed in the previous section that RQMC is more efficient in low dimensions. This allows for a better analysis of the choice of the sequence. For visibility, we  only show the results for the two test functions $\phi_1 : x \mapsto x_1$  and $\phi_2 : x \mapsto \sin(10 \cdot x_1) \cdot x_1^4$. For each test function, we take 100 samples of $2^{17}$ points from the different sequences/transforms for a common flow. The results are shown in Figure \ref{fig:dualmoon:seq}.

The Figure \ref{fig:dualmoon:seq} plots, in log-log scale, the mean of the absolute value of the error as a function of the sample size. We see that the two graphs are similar, up to a a vertical shift, corresponding to a multiplicative constant in the converging rate. The error for MC follow a $O(n^{-1/2})$ convergence rate, as expected. For RQMC, the four methods give a similar error, with a convergence rate that accelerate with the sample size (the slope of the line is increasing). The best performing method is the Sobol' sequence with a Box-Muller transform. Overall, Sobol' sequences perform better than Halton sequences, and the Box-Muller transform is better than the inverse transform. Worth noticing that for the Sobol' sequence, the error is significantly lower when the sample size is a power of 2. This is coherent with other results on Sobol' sequences \parencite{owenDroppingFirstSobol2022}. 

\subsubsection{Results for Markov chains methods}
\begin{figure}[h]
  \vspace{.25in}
  \centerline{\includegraphics[width=1.1\linewidth]{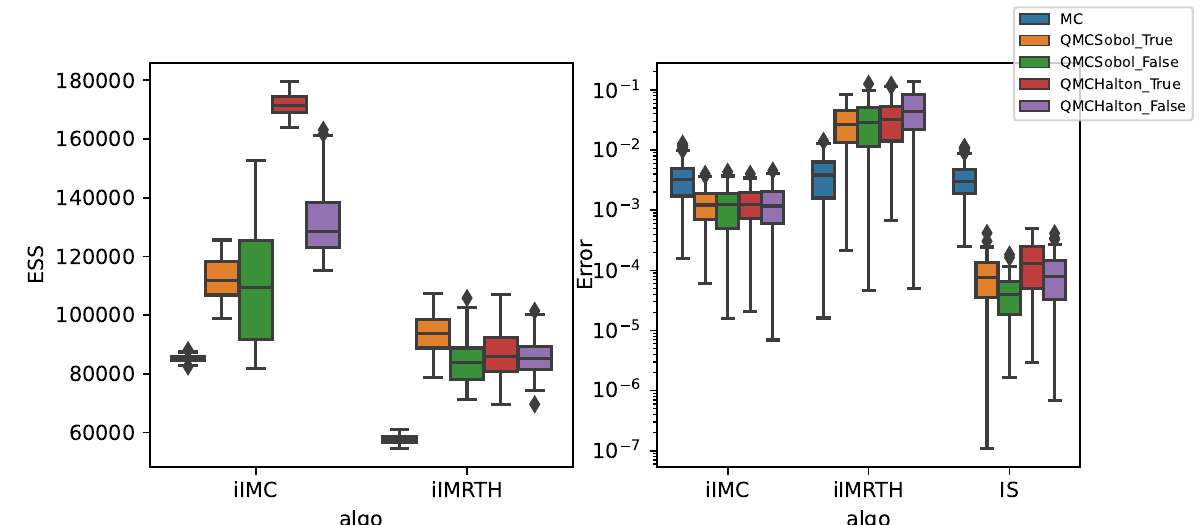}}
  \vspace{.3in}
  \caption{Results for the iMRTH and iIMC methods for the dualmoon target in dimension 2. (Left) the effective sample size for the different sequences. (Right) the absolute error of the estimate for the different sequences, as well as the error of the importance sampling estimator.}
  \label{fig:dualmoon:markov}
  \end{figure}

For the last numerical experiment we try two Markov-chain-based methods : the independent MRTH (iMRTH) and the independent IMC (iIMC) defined in section \ref{seq:montecarlo}. Both methods use  i.i.d. proposals from the instrumental distribution $\rho$  to construct a Markov chain that admits $\pi$ as invariant measure. As for importance sampling, we replace the $i.i.d.$ sample from $\rho$ for an RQMC sample using different sequences. The same data are used as in the previous section, with the same flow. All the analysis is done for the mean of the first component. The results are shown in Figure \ref{fig:dualmoon:markov}.

The effective sample size is a common tool for Markov chain analysis and is computed using the autocorrelations the chain \parencite{robertMonteCarloStatistical2010}. The higher the value, the better the chain. The plot on the left of Figure \ref{fig:dualmoon:markov} shows the effective sample size for the different sequences. For the iIMC, it shows a great change of efficiency between the different sequences and transforms, and always an increase of efficiency when using RQMC instead of MC. However, looking at the absolute error of the estimate on the right of the Figure, the four different RQMC estimates show very similar results. This behavior and the fact that the effective sample size is not a good indicator of the quality of the estimate in case of quasi-Monte Carlo is not surprising. Computing the ESS requires the proposals to be independent from one another. This is not the case for RQMC, as the points are correlated (or even anticorreleted).  The comparison between the iIMC and IS is in favour of IS: while the performances is similar with MC, IS is much more efficient with RQMC. 

For the iMRTH, the results are quite different. The effective sample size increases with RQMC but the absolute error of the estimate is worse than with  MC. In particular, the mean of the iMRTH chained is heavily biased for the four RQMC sequences, which is not the case for iIMC. This can be explained as the two methods present two very different behavior. For an iIMC chain, the number of replications of a point depends only on the point itself, not on the past or the future. For a iMRTH chain, the number of replications will depend on the following proposals from the flow. Therefore, as the RQMC points are not independent, the whole process is biased. This result is coherent with previous results on the use of RQMC for Markov chain Monte Carlo \parencite{owenQuasiMonteCarloMetropolis2005,liuLangevinQuasiMonteCarlo2023}. In these papers, the authors use a different kind of RQMC sequences, called completely uniformly distributed (CUD) sequences to make RQMC works with MCMC. 
\section{Conclusion}

In conclusion, we proposed to replace the classic i.i.d. sample from Monte Carlo for generating a trained normalizing flow with a randomized quasi-Monte Carlo sequence. We showed in several experiments that this can lead to a significant variance reduction and a better convergence rate. The main drawback is the scaling with the dimension, as the efficiency of RQMC is rapidly decreasing with the dimension.

The choice of the RQMC sequence is important, and the best choice in our experiments was a scrambled Sobol' sequence with a length $2^k$. The Box-Muller transform seems to be better than the inverse transform but this needs to be confirmed in further experiments. As expected from previous works, applying RQMC to MCMC methods can lead to biased results, in particular for the independent Metropolis-Rosenbluth-Teller-Hastings method. The independent importance markov chain method seems more robust to the use of RQMC, but still less efficient than importance sampling.

Future works could explore the use RQMC for training normalizing flows and not only the exploitation of the trained flow. This could be made easier if deep learning frameworks implement RQMC sequences. 

\section*{Acknowledgements}
The author was supported by a grant from Région Ile-de-France.

\newpage






 \printbibliography

\end{document}